# Relationships of Earthquake Moment Magnitude with Body and Surface Waves Using MCMC Approach

Zahra Firoozeh[1]*, Somayeh Taran[2]†, Nasila Asaadi[1]‡, and Hossein Safari[1]§

[1]Department of Physics, Faculty of Science, University of Zanjan, University Blvd., Zanjan, 45371-38791, Zanjan, Iran
[2]Department of Theoretical Physics and Astrophysics, Faculty of Physics, University of Tabriz, P.O. Box 51666-16471, Tabriz, Iran
October 13, 2024

## 1 Abstract

Due to the saturation of the body ($m_b$) and surface ($M_S$) earthquake magnitudes, the moment magnitude ($M_W$) is a more convenient parameter for representing earthquake energies. We use the HRVD data, including 18,569 earthquakes recorded from 1976 January 1 to 2010 December 31, for five regions on the Earth (**Asia**-Oceania, Europe, Africa, North America, and South America). Applying the error propagation technique on the moment tensor, we estimate the standard error for $M_W$ showing a distribution that deviates from Gaussian errors which recommends using the Monte Carlo Markov chain (MCMC) to obtain model parameters instead of the least square approach. We investigate the linear relationship between the body and surface waves in small ($\leq 6.1$) and large ($>6.1$) magnitudes via MCMC. The slope varies from 0.811 to 1.565 ($m_b - M_W$) and 0.566 to 0.971 ($M_S - M_W$) across five regions. For 18,569 global earthquakes, we obtained the slope for $m_b \leq 6.1$ ($m_b >6.1$) 0.863 (1.374), while we find the slope for $M_S \leq 6.1$ ($M_S > 6.1$) 0.581 (0.921). These transition relationships of the magnitude moments are useful for increasing earthquake monitoring capacity.

---
*E-mail: zahra.firoozeh@znu.ac.ir
†E-mail: taran@tabrizu.ac.ir
‡E-mail: nazilaasaadi@znu.ac.ir
§E-mail: safari@znu.ac.ir

# 2 Introduction

One of the most important goals of seismology is to try to understand the physical process of this phenomenon. For this purpose, it would be possible to analyze all the details recorded in an event, which is not practicable due to the time-consuming and impossibility of recording all the details of an earthquake (Kanamori, 1983; Mousavi and Beroza, 2020). Historically, the most well-known measure of the earthquake scale is the moment magnitude (Yoder et al., 2012; Shearer, 2019). Earthquake magnitude scales usually correspond to the largest amplitude recorded in a seismogram, which is the most straightforward measurement. The moment scale is very famous for earthquake studies (Di Giacomo et al., 2015). The seismological centers worldwide use several magnitude scales. Also, sometimes some of these centers record one or two magnitudes of earthquakes (such as the USGS center) because of limitations or difficulty in measuring their size. Understanding earthquakes' structure, physical interpretations, and predictions requires using different magnitude scales to represent earthquakes' released energy (intensity) (Okal, 2019). It is necessary to discover scaling relationships to transform various magnitude scales, such as body-wave magnitude ($m_b$) and surface-wave magnitude ($M_S$), into a homogeneous magnitude like moment magnitude ($M_W$) (Jordan et al., 2011).
Scaling relationships can be used to calculate additional event parameters for comprehensive earthquake studies and maintain consistency between existing and new data. Furthermore, an accurate relationship can estimate the maximum magnitude within a region (Vakov, 1996; Lunina and Gladkov, 2015).
Numerous studies have been performed to provide the empirical relationships between magnitude scales using various methods in different regions (Stromeyer et al., 2004; KADİRİOĞLU and Kartal, 2016; Herak, 2020). These standard methods are the ordinary least square (OLS) regression (Draper, 1981), general orthogonal regression (GOR) (Fuller, 1987), and other similar orthogonal regression techniques (Stromeyer et al., 2004; Krystek and Anton, 2007). The OLS regression methods are used to convert the teleseismic magnitudes ($M_S$ and $m_b$) to moment magnitude ($M_W$) via the assumption that the uncertainty of the inde-pendent variable is negligible (Chiaruttini and Siro, 1981; Heaton et al., 1986; Johnston, 1996; Giardini et al., 1997; Gasperini et al., 2000; Scordilis, 2006). Due to the bias in the Gutenberg Richter frequency-magnitude distribution, Castellaro and Bormann (2007) showed that the OLS regression method is inappropriate for converting magnitude scales.

So, they proposed the GOR method, which is where the error ratio on the independent and dependent variables is available. Scientists performed different uncertainties for dependent and independent variables in the standard GOR and similar orthogonal regression techniques (Bormann et al., 2007; Grünthal et al., 2009; Ristau, 2009; Deniz and Yucemen, 2010; Bethmann and Luhmann, 2010; Das et al., 2011; Baruah et al., 2012; Das et al., 2012; Gasperini et al., 2012; Grünthal and Wahlström, 2012; Das et al., 2014; Pujol, 2016; Goitom et al., 2017; Nath et al., 2017).

Lolli and Gasperini (2012) demonstrated that general orthogonal linear regression methods, chi-square regression, general orthogonal regression, and weighted least-square approach yield nearly identical regression coefficients and uncertainties.

The common assumption of the above three methods is that the **error** variance ratio ($\eta$) denotes the ratio of the **error** variance of dependent and independent variables.

In most studies, the conventional GOR obtains the true variable ($Y_t$) by substituting the independent variable ($X_{obs}$) in the GOR relation. Carroll and Ruppert (1996) and other researchers (e.g., Stefanski (2000); Carroll et al. (2006); Das et al. (2018)) reported a significant overestimation problem in the GOR method. The GOR method ignores the error equation and estimates the error ratio on the available independent and dependent variables.

Das et al. (2018) suggested the proposed GOR method to reduce the bias in the dependent variable. In the proposed GOR, $Y_t$ cannot be obtained by substituting the $X_{obs}$ in the GOR relation, so they expressed the linear relation between $X_{obs}$ and $Y_t$ values. Different methods used for conversion problems produce bias in estimating $Y_t$ because they cannot account for errors in the dependent and independent variables.

The normal distribution of errors and independent variables is the main assumption for modeling data based on least squares fitting, which is the outcome of maximum likelihood estimation (MLE) (Pan et al., 2002; Suh, 2015). Maximizing a likelihood function makes the probability of the data observed in the hypothetical model the most probable (Chambers et al., 2012; Di Giacomo et al., 2015; Rossi, 2018). For the differentiable likelihood function, the derivative test is one of the essential ways to determine (explicitly or numerically) the maxima (e.g.,

Press et al. (1992); Gui et al. (2023)). In the Bayesian approach, the optimal model determined by the MLE of maximal posterior assumed the prior uniform distribution of parameters (Ward and Ahlquist, 2018). Traditional methods like MLE may not be appropriate for determining model parameters when dealing with non-Gaussian distributions. In the context of non-Gaussian distribution of uncertainties, the Markov Chain Monte Carlo (MCMC) method is a suitable approach to estimate the parameters of complex models and quantify uncertainty in a Bayesian framework (Delle Monache et al., 2008; Chen et al., 2023). The MCMC methods, which include Monte Carlo random walking methods, are a set of algorithms used to sample probability distributions based on building a markup chain with the right features (Zheng and Han, 2016). The chain mode is an example of a suitable distribution after many steps. The quality of this sample increases with the increasing number of steps (Kang, 2005). The Metropolis-Hastings algorithm is among the many algorithms for building chains (Calderhead, 2014).

The maximum posterior probability estimate approximates an unknown quantity equivalent to the state of the posterior distribution (Farhang et al., 2018).

When direct sampling of a series of random samples is complex, the Metropolis-Hastings algorithm can generate a sequence of samples from a probability distribution (Berg, 2004; Dunson and Johndrow, 2020). The mentioned sequence sampling is useful for determining the distribution and calculating an integral (Chib and Greenberg, 1995; Jusup et al., 2022).

Applying Metropolis-Hastings and other MCMC algorithms is recommended when dealing with many dimensions (Gilks et al., 1995; Lee and Wagenmakers, 2014). In computational physics (Kasim et al., 2019), computational linguistics (Robert et al., 2004; Gill, 2008), and computational biology (Gupta and Rawlings, 2014) can easily see the traces of the MCMC methods for calculating multi-dimensional integer numerical approximations. The Bayesian statistics used the MCMC methods to effectively integrate numerous unknown parameters in the complex hierarchical model computation (Delle Monache et al., 2008; Turek et al., 2016).

Here, we obtain the standard errors of the earthquake moment magnitude using the error propagation of moment tensors for the Harvard data center. Since 1976, the Harvard Center has reported earthquake magnitudes in the scale of $m_b$, $M_S$, and $M_W$ to characterize seismic events. These scales are helpful for earthquake monitoring and research because they provide comprehensive information about the size and intensity of the earthquake. The $M_L$ scale (local magnitude) is based on the maximum amplitude of a seismogram recorded on a Wood-Anderson torsion seismograph. Data centers do not as commonly report it for larger events. The $M_L$ values are calculated using modern instrumentation with appropriate adjustments, but its authoritative use is typically limited to smaller events, particularly those with magnitudes less than 4.0 (Deichmann, 2017, 2018; Augliera, 2022). Therefore, we focus on investigating the relationship between $M_W$ with $m_b$ and $M_S$, as recorded by Harvard. The non-Gaussian distribution of $M_W$ magnitude errors allows us to apply the MCMC approach for a set of observations of $M_W$ and $m_b$ (or $M_S$) for each Earth's five regions (**Asia**-Oceania, Europe, Africa, North America, and South America) and global Earth to determine the linear transition relationship's parameters (slope and intercept). The MCMC method is an approach to infer the parameters of a model given observational data in Bayesian statistics. We define the likelihood function for the linear transition relationship of magnitudes including standard errors. Then, we determine the slope and intercept in the MCMC sampling (Metropolis-Hastings) on the posterior parameters using the posterior distribution including a prior distribution and likelihood. Also, the upper and lower limits of errors for parameters are calculated.

## 3 Data Set

In this study, we use the 18,569 earthquakes of the HRVD, recorded from 1976 January 1 to 2010 December 31. For this data set, we use $m_b$ (body wave magnitude), $M_S$ (surface wave magnitude), and $M_W$ (moment magnitude) ranging from 4.1 to 8.9, 2.4 to 8.9, and 4.3 to 9, respectively. Hanks and Kanamori (1979) showed that the earthquake moment magnitudes, e.g., $M_S$ and $m_b$, are statistically correlated, have regional differences, and depend on the earthquake's source mechanism (Chen et al., 2001). The reports from the Harvard data center summarize the number of observations relevant for assessing the past and current quality of data recorded at a station within the Global Seismographic Network (Arrais et al., 2022). Some issues are related to errors in the available descriptions of station parameters, such as sensor orientation, response functions, and polarities. Updating station parameters can theoretically correct errors in station metadata. In practice, this may be challenging in certain situations due to a lack of knowledge or an inability to determine the appropriate parameters. The Harvard data center evaluates well-performing stations to establish their high quality and to identify installation and operational procedures that could be replicated at other stations. For example, the SSE station (Shanghai) is situated on the outskirts of Shanghai, China. It is well-positioned to provide global coverage for earthquake monitoring, nuclear monitoring, and Earth structure studies. The closest GSN station is TATO-IU (Taipei) in Taiwan, approximately 700 km south of SSE (Iwata, 2008; Chang et al., 2016). Also, the Harvard data center considers the spatial and temporal considerations of the data on a process (Ekström and Nettles, 1997; Huang et al., 1997; Chen et al., 2001).

One approach to evaluating data quality entails systematically comparing recorded long-period waveforms with synthetic seismograms simulated for confirmed seismic events. This assessment methodology adheres to the protocol delineated by Ekstrom (2006). Data collection involves seismic data from both the LH and VH channels of the STS-1 and STS-2 sensors. Synthetic waveforms corresponding to earthquakes listed in the Global CMT catalog (Dziewonski et al., 1981; Ekström et al., 2005) with a moment magnitude ($M_W$) of 6.5 or higher are computed for analysis. Calculation of correlation coefficients and optimal scaling factors is performed across three standard CMT data types: body waves ($B$) spanning periods of 50 to 150 seconds, mantle waves ($M$) spanning periods of 125 to 350 seconds, and surface waves ($S$) spanning periods of 50 to 150 seconds. This discussion primarily examines findings associated with body and mantle waves. The scaling factor is determined solely for waveforms exhibiting a correlation coefficient of 0.75 or greater, signifying the value by which the synthetic seismogram should be adjusted to better align with the observed seismogram. Annual median scaling factor values are derived when four or more individual event scaling estimates are available for the respective year. Notably, negative correlations indicate the presence of reversed components (Ekström et al., 2010).

Figure 1 represents the Earth seismic map including five different regions, **Asia**-Oceania (yellow), Europe (pink), Africa (green), North America (blue), and South America (orange). The earthquake moment magnitude ranges 4.3 to 9.0, indicated with circle markers and different colors. We study the transition relationship between different types of magnitudes ($m_b$, $M_S$, and $M_W$) in each area.

The bulletin of the HRVD (https://www.globalcmt.org/CMTsearch.html) data center provides information on the magnitude of earthquakes. We need the uncertainties for magnitudes (Dziewonski et al., 1981; Ekström et al., 2012). However, in the received data from the HRVD center, there are data tensor moments without the uncertainties of $M_W$. The moment tensor is a mathematical description of the forces causing deformation and rupture at the earthquake source, producing seismic waves that reveal the nature of the fault (Ammon et al., 2021). So, to obtain the uncertainty for $M_W$, we use the following relation of moment tensor

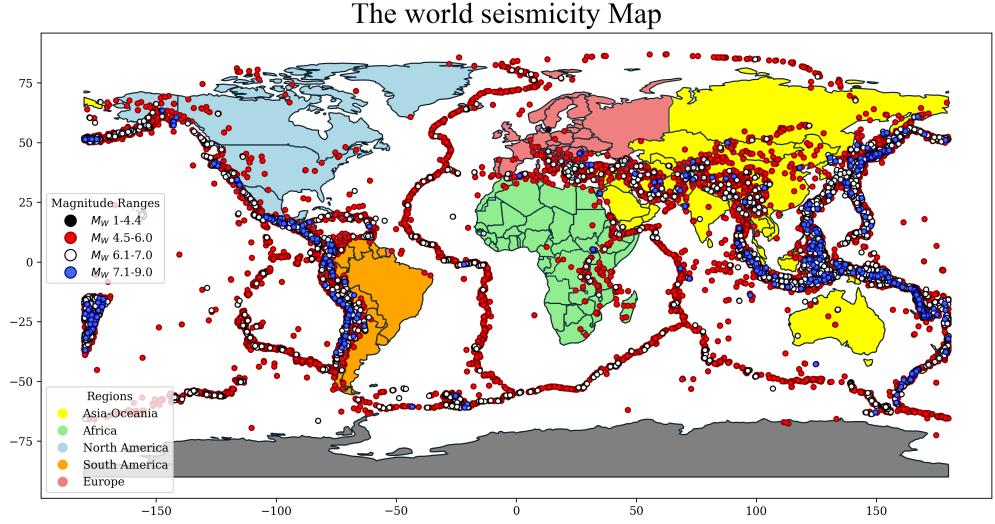

Figure 1: Earth seismic map divided into five regions **Asia**-Oceania (yellow), Europe (pink), Africa (green), North America (blue), and South America (orange). The black, red, white, and blue circles indicate the earthquake moment magnitude of 1-4.4, 4.5-6, 6.1-7, 7.1-9, respectively.

(**M**) (Shearer, 2019),

$$\mathbf{M} = \begin{bmatrix} M_{11} & M_{21} & M_{31} \\ M_{12} & M_{22} & M_{32} \\ M_{13} & M_{23} & M_{33} \end{bmatrix}, \tag{1}$$

where $M_{ij}$ is a pair of antonym forces in $i$ ($= 1, \cdots, 3$) of Cartesian coordinates that separated in $j$ ($= 1, \cdots, 3$) direction. According to angular momentum conservation, the **M** tensor should be symmetric (e.g., $M_{ij} = M_{ji}$). Thus, **M** consists of six independent elements. Table 1 presents a sample of HRVD data that includes the six moment-tensor elements $M_{rr}$, $M_{tt}$, $M_{pp}$, $M_{rt}$, $M_{rp}$, and $M_{tp}$ and their standard errors. The index $r$, $t$, and $p$ denote the upward, southward, and eastward, respectively. Therefore, we can first obtain $M_0$ from $M_{ij}$ elements (the scalar seismic moment is $M_0$), that the moment tensor presents an overview of the forces created inside the Earth which can act at a point in an elastic medium (Shearer, 2019),

$$\mathbf{M} = \begin{bmatrix} 0 & M_0 & 0 \\ M_0 & 0 & 0 \\ 0 & 0 & 0 \end{bmatrix}. \qquad (2)$$

The unit of $M_0$ is N-m like a force pair and can commonly be calculated from any moment tensor as Shearer (2019),

$$M_0 = \frac{1}{\sqrt{2}} \left( \sum_{ij} M_{ij}^2 \right)^{1/2}. \qquad (3)$$

Aki (1965) stated that this scalar seismic moment is the most basic measure of earthquake strength.

Table 1: The full format used to store and distribute the Global Centroid-Moment-Tensor (CMT) on the Harvard CMT catalog, which contains the six moment-tensor elements $M_{rr}$, $M_{tt}$, $M_{pp}$, $M_{rt}$, $M_{rp}$, and $M_{tp}$ and their standard errors. The $r$, $t$, and $p$ denote the north, south, and east directions, respectively.

|       | $M_{rr}$ | $M_{tt}$ | $M_{pp}$ | $M_{rt}$ | $M_{rp}$ | $M_{tp}$ |
|-------|----------|----------|----------|----------|----------|----------|
| CMT   | 1.100    | -0.300   | -0.800   | 1.050    | 1.240    | -0.560   |
| Error | 0.020    | 0.020    | 0.020    | 0.070    | 0.080    | 0.020    |

The moment magnitude is measured by the components of CMT from the HRVD observations, including the uncertainties likely related to the components' systematic errors (estimated standard errors). Therefore, we can estimate the systematic errors for moment magnitude using error propagation. We determine the errors for HRVD data via the error propagation that is applied to the moment tensor (Equation 3) as follows,

$$\delta M_o = \sqrt{\sum_{ij} \left( \frac{\partial M_0}{\partial M_{ij}} \right)^2 \sigma_{ij}^2} = \frac{1}{M_0} \sqrt{\sum_{ij} M_{ij}^2 \sigma_{ij}^2}, \qquad (4)$$

where $\sigma_{ij}$ is the tensor of the expected standard deviation of the $ij$th residual related to random measurement error.

The moment magnitude $M_W$ is given by,

$$M_W = \frac{2}{3} (\log M_0 - 16.1), \qquad (5)$$

and the uncertainty for $M_W$ is defined by,

$$\delta M_W = \frac{2}{3} \frac{\delta M_0}{M_0}. \tag{6}$$

Since measuring stations have inevitable measurement errors, Equation (6) gives a fraction of the standard error for $M_W$.

# 4 Methods

Several methods were investigated to fit a straight line for data. Standard weighted least-squares (LS) fitting and maximum likelihood estimation (MLE) are suitable choices for the data points with negligible uncertainties (Akritas and Bershady, 1996; Van Herk and Dröge, 1997; Jaynes, 2003; Cohen et al., 2013). The MLE estimates the parameters of a model fitted to the frequency distribution by maximizing a likelihood function (Moradhaseli et al., 2021). Our data encounter heterogeneous and arbitrary uncertainties and situations with significant outliers, unknown uncertainties, and inherent scatter in the linear fitting. It is worth mentioning that having a production model for the data is essential. In the presence of a generative model, the subsequent fitting is non-arbitrary because the model allows direct computation of likelihood parameters (in the maximum likelihood estimation) or the posterior probability distribution. Constructing a posterior probability distribution becomes necessary for marginalized nuisance parameters. So, we apply the MCMC approach to obtain the parameters of linear transition relationships between the moment magnitude $M_W$ containing its non-Gaussian errors and body wave magnitude $m_b$ (or surface wave magnitude $M_S$).

For a data set $(x_n, y_n)$ with $n = 1, \cdots, N$ and their non-Gaussian variances $\sigma_n$, a linear model ($y_n = \alpha x_n + \beta$) with a slope $\alpha$ and an intercept $\beta$, the logarithm of the likelihood function is given by (Hogg et al., 2010; Murphy, 2012),

$$\ln p(y \mid x, \sigma, \alpha, \beta, f) = -\frac{1}{2} \sum_n [\frac{(y_n - \alpha x_n - \beta)^2}{s_n^2}] + \ln(2\pi s_n^2), \tag{7}$$

where $s_n^2 = \sigma_n^2 + f^2(\alpha x_n + \beta)^2$. The parameter $f$ is some fractional amount that presents the underestimation of the variance. Since $\ln p(y \mid x, \sigma, \alpha, \beta, f)$ is an increasing function, the maximum of the likelihood and log-likelihood coincide. We must consider that the critical point in an optimization method for model parameters must be a maximum value. For this purpose, we set the partial derivatives equal to zero,

$$\frac{\partial \ln p}{\partial \alpha} = \sum_n [\frac{x_n(y_n - \alpha x_n - \beta)}{s_n^2} + \frac{f^2(y_n - \alpha x_n - \beta)^2(\alpha x_n^2 + \beta)}{s_n^4}] + \frac{f^2(\alpha x_n^2 + \beta x_n)}{s_n^2}, \tag{8}$$

$$\frac{\partial \ln p}{\partial \beta} = \sum_n [\frac{(y_n - \alpha x_n - \beta)}{s_n^2} + \frac{f^2(y_n - \alpha x_n - \beta)^2(\beta + \alpha x_n)}{s_n^4}] + \frac{f^2(\alpha x_n + \beta)}{s_n^2}, \tag{9}$$

$$\frac{\partial \ln p}{\partial f} = \sum_n [\frac{f(\alpha x_n + \beta)^2(y_n - \alpha x_n - \beta)^2}{s_n^4}] + 2\frac{f(\alpha x_n + \beta)^2}{s_n^2}. \tag{10}$$

One may solve these equations to obtain the fit parameters $\alpha$, $\beta$, and $f$.

### 4.1 MCMC

The MCMC procedure generates a systematic random sampling from a known probability distribution (Spall, 2003). In practice, a set of chains sufficiently separated from each other is usually created for collecting points (Foreman-Mackey et al., 2013). These chains are explored using the Metropolis-Hastings algorithm, which is a crucial part of the MCMC process. In this algorithm, a new state (or point) is proposed based on the current state, and this new state is accepted or rejected according to the Metropolis criterion. This criterion ensures that states with higher probability are more likely to be accepted, guiding the chains to regions of the parameter space where the integrals have higher values. This method allows the chains to move through the parameter space in a way that preferentially explores regions with higher posterior probability, efficiently sampling from the distribution.

While the random integral samples used in a typical Monte Carlo integration are statistically independent, the samples used in the MCMC method are auto-correlated. To compute the mean value error, it is necessary to use the Markov chain central limit theorem to determine the correlation between the samples (e.g., Šukys and Kattwinkel (2017), Figure 2 therein). The Markov property is essential here, as it ensures that each sample in the chain only depends on the previous one, allowing the chain to converge to the target distribution eventually. This property also underpins the algorithm's efficiency in exploring complex, high-dimensional spaces where traditional sampling methods might fail.

When uncertainties (like standard errors of $M_W$) are non-Gaussian, we can marginalize some disturbing parameters to obtain the best model parameters. To best estimate the posterior probability function (the distribution of parameters matches our data set), we employ the MCMC method to propagate error parameters. The posterior probability function is given by:

$$p(\alpha, \beta, f \mid x, y, \sigma) \propto p(\alpha, \beta, f) p(y \mid x, \sigma, \alpha, \beta, f), \tag{11}$$

where $p(y \mid x, \sigma, \alpha, \beta, f)$ is the likelihood function (Equation 7), and $p(\alpha, \beta, f)$ is the prior function. This prior function incorporates any prior knowledge about the parameters, including results from other similar experiments, admissible physical limits, and other information. The MCMC method requires introducing a probability distribution for sampling the model parameters (Hogg et al., 2010; Robert et al., 2018). Next, we must select priors on the model parameters $\alpha, \beta$, and $f$. For example, we consider a uniform distribution for a conservative prior to $\alpha$ as:

$$p(\alpha) = \begin{cases} 1/2, & \text{if } 0 \leq \alpha \leq 2 \\ 0, & \text{otherwise} \end{cases}. \tag{12}$$

Then, we assume the uniform distribution for $\beta$:

$$p(\beta) = \begin{cases} 1/4, & \text{if } -2 < \beta < 2 \\ 0, & \text{otherwise} \end{cases}. \tag{13}$$

We then combine the log-prior with the log-likelihood function. For this purpose, we use the emcee package that includes these steps (Hogg et al., 2010). Foreman-Mackey et al. (2013) proposed emcee as a stable, pure-Python implementation of the affine-invariant ensemble sampler for MCMC. The emcee package is particularly effective because it adapts to the problem at hand by measuring the autocorrelation time, ensuring that the algorithm performs better than traditional MCMC methods by efficiently exploring the parameter space. Using a primary guess, the emcee package initializes walkers in a small Gaussian ball around the maximum likelihood result and then runs several steps of MCMC. We use the slope and intercept estimated by the least square approach as the primary guess via the MCMC (Metropolis-Hasting) sampling to achieve fast convergence in the program and find the best parameter estimation. The walkers move as follows: first, in small distributions around the maximum probability values, and next, they have a kind of rapid wandering to explore the complete posterior distribution. We observe that in about 50 steps, the code samples converge to the best parameters for our data sample.

Using the corner plots, we can interpret the MCMC outputs. A corner plot shows all the projections of the posterior probability distributions of our parameters. The corner plots are advantageous for displaying all covariances between parameters. A method for generating a marginal distribution for a parameter or set of parameters using MCMC chain results is projecting the samples onto that screen. The other way is to project our results in the data space observed by a diagnostic chart. For this purpose, we can select several data samples from the chain and plot them above the data points (Draper, 2008).

# 5 Results

Here, we use earthquake information provided by HRVD for 18,569 earthquakes. To homogenize earthquake magnitudes, we convert $m_b$ and $M_S$ to $M_W$. Moreover, the relation of magnitudes of different types for a particular event carries essential information about the nature and physics of this event because it reflects the portion of seismic energy radiated at different frequencies.

We divided the Earth's surface into five regions to study the relationship between magnitudes. Figure 1 displays the seismic map for the five regions and the moment magnitude of earthquakes. 4049, 2647, 4314, 4381, and 3178 earthquakes were analyzed for the **Asia**-Oceania, Europe, Africa, North America, and South America, respectively. The geopolitical boundaries and practical considerations related to data availability, and collection are reflected in the division (Doglioni et al., 1999; Stein and Wysession, 2009; Shearer, 2019). Our analysis is aligned with existing frameworks used by international agencies (e.g., IASPEI and GSN) by grouping countries and territories into these broader regions, making it easier to compare and apply our findings. Using data distribution and regional representation, we can ensure a more balanced distribution of data across different parts of the world by dividing it. While we acknowledge that no region is perfectly homogeneous, the division reduces potential biases that may result from the over-representation of data from more seismically active areas, such as the Pacific Ring of Fire, in a global analysis (Doglioni et al., 1999). We aim to capture and account for regional variations in seismic behavior by examining these regions separately. Many studies investigated earthquake magnitude scale relationships locally, and this division allows us to compare our results with previous studies (Di Giacomo et al., 2015; KADİRİOĞLU and Kartal, 2016; Kumar et al., 2020; Boudebouda

et al., 2024).

First, applying the error propagation method (Equations 4 - 6) and observed moment tensors, we estimate the standard errors $\delta M_W$ (Section 3). The Probability density function (PDF) for standard errors of moment magnitudes shows a deviation from Gaussian distribution (Figure 2). Figure 2 shows the frequency-size distribution of $M_W$ standard errors for the HRVD data set in the **Asia**-Oceania, Europe, Africa, North America, and South America, respectively. As the figures show, the distributions follow a non-Gaussian behavior.

Fitting methods like least square and maximum likelihood estimations are not commonly recommended for data with non-Gaussian distribution errors. Second, we used the MCMC approach for $M_S$ or $m_b$, $M_W$, and $\delta M_W$ to obtain the slope ($\alpha$) and intercept ($\beta$). Also, the upper and lower limits of standard errors for $\alpha$ and $\beta$ are given by the MCMC approach (Section 4.1). For the goodness of fit, we measure $R$−squared and $p$−value for each fit.

Figure 3 shows the sample of posterior parameters (Equations 11, 12, and 13) slope ($\alpha$), intercept ($\beta$), and fraction (log $f$) variations in terms of step numbers in the chain for **Asia**-Oceania earthquake data. The slope and intercept indicate the linear correlation between $M_W$ and $m_b (\leq 6.1)$. The model parameters show almost oscillating behavior (fluctuating around the average values) due to random data sampling by the MCMC approach. The similar behavior for transition relation parameters of $m_b > 6.1$ and another four Earth regions were observed.

Figure 4 displays the correlation and one-dimension marginal distribution of the posterior probability distributions of $M_W$ and $m_b (\leq 6.1)$ relationship parameters ($\alpha$, $\beta$, and log($f$)) using a corner plot diagram with their true values of the MCMC method in blue squares. In the context of MCMC methods, the corner plot is a visualization tool used to explore the posterior distribution of model parameters. When conducting Bayesian inference with MCMC, we are interested in estimating the posterior distribution of parameters given observed data. The corner plot displays a pairwise scatter plot for each combination of parameters, histograms along the diagonal showing the marginal distributions of individual parameters, and often, contours or shading indicating regions of high posterior density (Spitoni et al., 2020). We observe similar behavior for corner plots of model parameters for $m_b > 6.1$ and the relation for $M_W$ and $M_S$ in five Earth regions.

This visualization helps us to understand the relationships between parameters ($m_b$ versus $M_W$ and $M_S$ versus $M_W$), identify correlations, assess convergence, and diagnose potential problems with the model or sampling procedure. By examining the corner plot, we can gain valuable insights into the behavior and uncertainty of our model's parameters ($\alpha$ and $\beta$), interpreting Bayesian analyses.

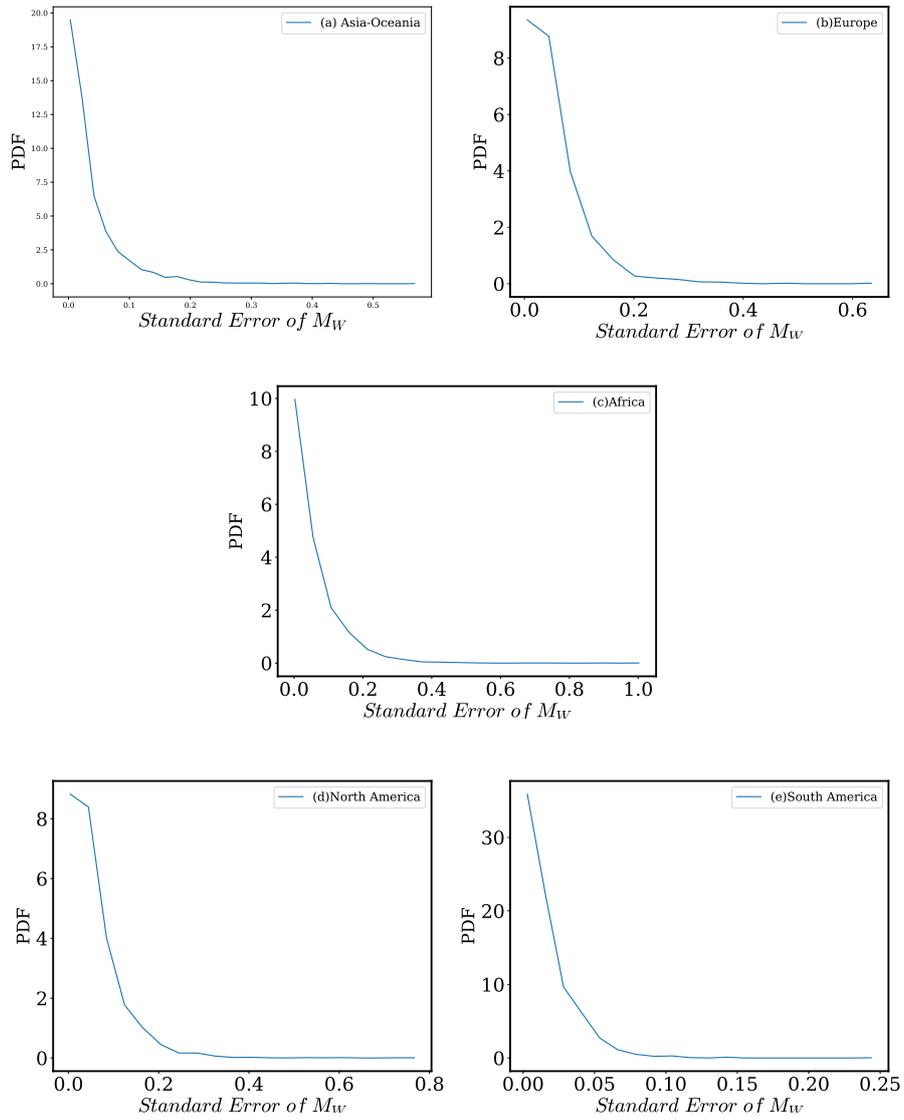

Figure 2: Distribution of $M_W$'s standard errors for 18,569, occurred earthquake points recorded from (a) **Asia**-Oceania, (b) Europe, (c) Africa, (d) North America, and (e) South America.

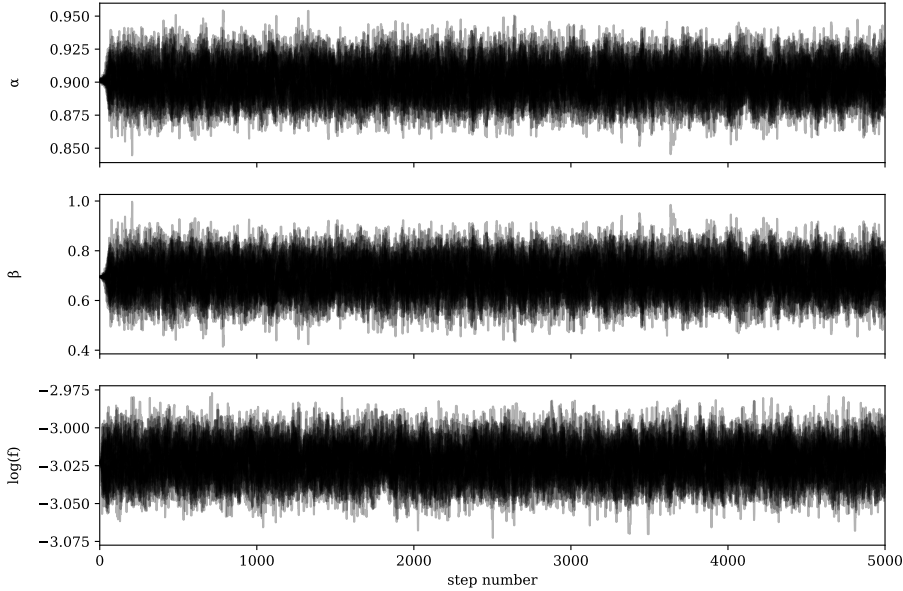

Figure 3: The value of parameters: slope ($\alpha$: top panel), intercept ($\beta$: middle panel), and a fraction ($\log f$: bottom panel) for each walker as a function of the number of steps in the chain for **Asia**-Oceania earthquakes. The slope and intercept parameters represent the linear relation of $M_W$ and $m_b (\leq 6.1)$. Because of random sampling in the MCMC algorithm, the model parameters ($\alpha$, $\beta$, and $\log f$) fluctuate around the true values.

The negative covariance between $\alpha$ and $\beta$ indicates their inverse relationship. The scatter plots of $\log(f)$ versus $\alpha$ and $\log(f)$ versus $\beta$ indicate no linear relationship. The posterior distributions are well-constrained and roughly Gaussian. Hence, the MCMC method chooses the average of the sampled parameters (blue squares in Figure 4) as the true value and the averages of maximum and minimum variations from the true value (Figure 3) as each parameter's upper and lower bound, respectively. Therefore, we obtained the slope ($\alpha = 0.923^{0.090}_{-0.062}$) and intercept ($\beta = 0.576^{0.475}_{-0.326}$) for sample data in which the super and sub-indexes indicate the parameters' upper and lower values, respectively. Please note that the asymmetry in the lower and upper error values in both slope and intercept might rooted in the non-Gaussian distribution of moment magnitude errors. The non-Gaussian errors for a data set are a limitation of using the least square approach to obtain model parameters. Indeed, the least square approach derived from the maximum likelihood estimation for the data set assumes the Gaussian distribution for standard errors (Mestav et al., 2019; Xiao et al., 2019).

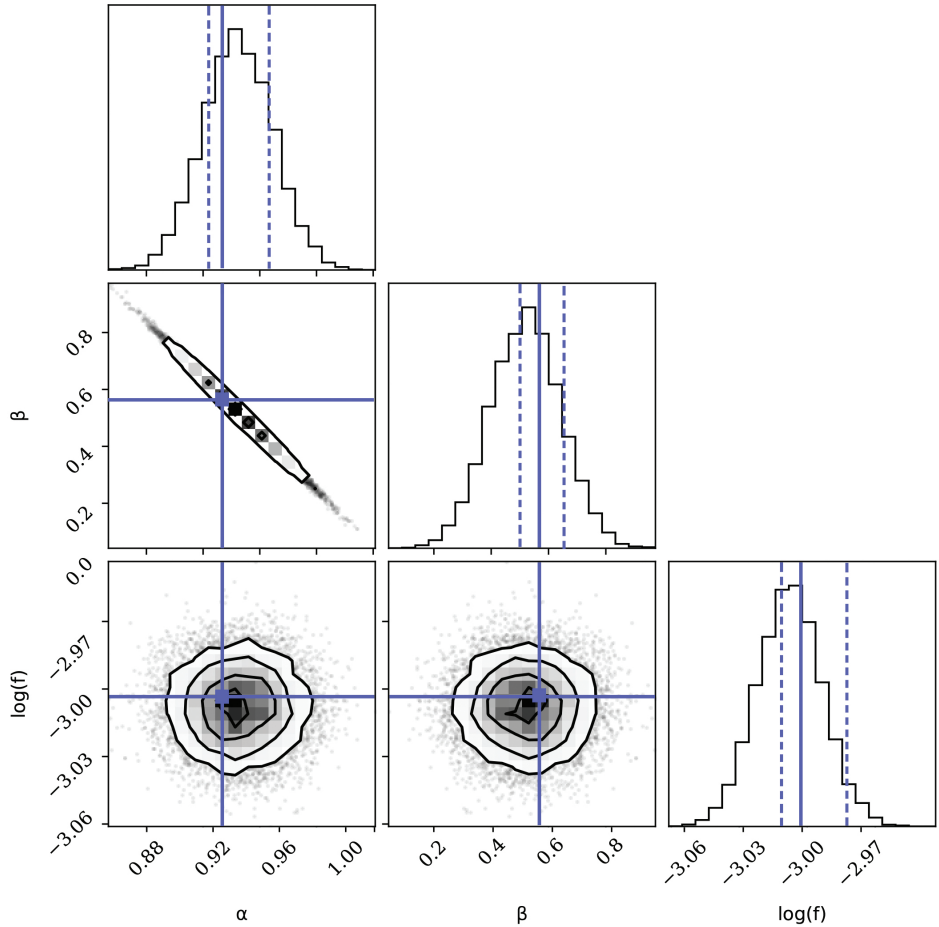

Figure 4: The corner plot of histograms and scatter plots of MCMC analysis for marginalized posterior parameters as the slope $\alpha$, intercept $\beta$, and (log-)fractional uncertainty $\log f$, with their true values in blue. The vertical dashed lines show the upper and lower bounds of each parameter (**Asia**-Oceania earthquakes).

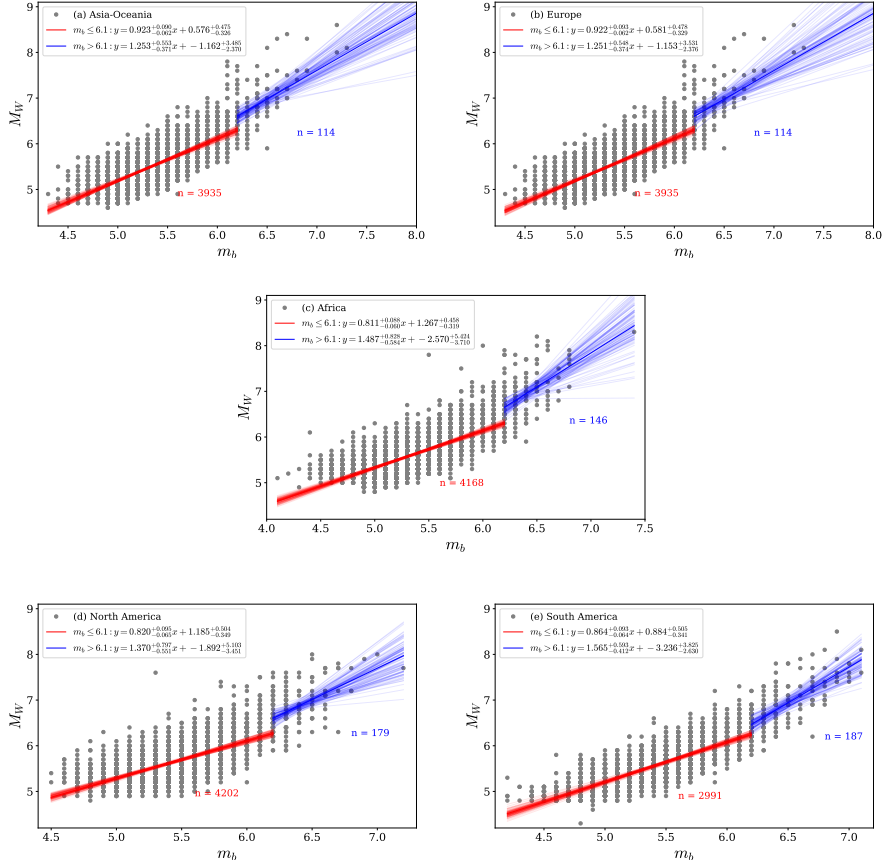

Figure 5: Transition relationship between $M_W$ and $m_b$ for $m_b \leq 6.1$ (red line) and $m_b > 6.1$ (blue lines) using MCMC method for earthquakes of (a) **Asia**-Oceania, (b) Europe, (c) Africa, (d) North America, and (e) South America, respectively. $n$ indicates the number of earthquakes for each region and magnitude range.

Figure 5 displays the MCMC results for the transition relationship of $M_W$ with $m_b \leq 6.1$ (red line) and $m_b > 6.1$ (blue lines) for five regions. Since the mean value of the first 5 seconds of short-period amplitude records for $m_b$, its saturation is expected. As a result, in strong earthquakes, the maximum amplitude of the earthquake wave may often be postponed and lead to an underestimation of magnitudes (Scordilis, 2006). Therefore, we obtained two linear models for body-wave magnitudes $\leq 6.1$ and > 6.1 which showing two different regimes of the relationship between $M_W$ and $m_b$. The slope varies in the range of $0.811 - 0.923$ for $m_b \leq 6.1$, while it changes in the range of $1.253 - 1.565$ for $m_b > 6.1$ in the five regions. Based on the MCMC approach, we obtain the relationship between $M_W$ and $m_b$, which is listed in Table 2. In a similar analysis, Scordilis (2006) reported a slope of about $0.85 \pm 0.04$ for a small magnitude regime of $m_b \leq 6.1$ for global earthquakes recorded during 1965-2003 by NEIC and ISC data centers. The range of slope true value obtained in the present study for small magnitude regime agrees agreement with Scordilis (2006). KADİRİOĞLU and Kartal (2016) obtained the slope $1.0319 \pm 0.025$ for the linear relationship of $M_W$ and $m_b$ for 489 earthquakes recorded by Harvard data center in Turkey and near surrounding from 1900 to 2012. Kumar et al. (2020) formulated earthquake magnitude conversion relations specifically for the Himalayan seismic belt, based on data from various seismological agencies such as the ISC, NEIC, and others. Their analysis utilized intra- and inter-magnitude regression techniques to enhance accuracy across scales. Notably, they found that certain magnitude relations, such as those reported by the Global CMT, ISC, and NEIC, yield small magnitudes with the relation $M_W = 0.93(\pm 0.07)m_b + 0.13(\pm 0.35)$.

Our findings indicate a slope of $\alpha = 0.923$ for the small magnitude range and $\alpha = 1.253$ for large magnitudes in the **Asia**-Oceania region, which aligns with the slopes obtained by KADİRİOĞLU and Kartal (2016) and Kumar et al. (2020). Gasperini et al. (2013) determined the slop in the range of $1.23 - 1.39$ for the linear relationship between $M_W$ and $m_b$ using different methods for their earthquake catalog including Global CMT (1976-2010), NEIC (1980-2009), European–Mediterranean Regional Centroid Moment Tensor (1997-2010), and the Regional MT catalog of the ETHZ (1996-2006). Their finding slopes are verified by the derived slope $\alpha = 0.922$ (for $m_b \leq 6.1$) - $1.251$ (for $m_b > 6.1$) applying the MCMC method in the present work for the Europe region. For Mexico region updated catalog from 1776 to 2017 including 712 earthquakes, Sawires et al. (2019) obtained the slope $1.35 \pm 0.15$ for the linear relationship of $M_W$ and $m_b$ (ranging from 4-7.1) which is a slight agreement with the slope $\alpha = 0.820 - 1.370$ for North America region in our study. Table 2 gives the different measures of the goodness of fits, such as $R$−squared and $p$−value for five regions on the Earth for two regimes of $m_b \leq 6.1$ and $m_b > 6.1$. The table outlines, that the $p$−values for all five Earth regions are significantly greater than the critical threshold value (0.05), indicating that the null hypothesis (the linear relationship between $M_W$ and $m_b$) is not rejected. In statistics, the $p$−value represents the probability of observing a test statistic as extreme or more extreme than the one calculated, assuming the null hypothesis is true. The $R$−square value ranges from 0.22 to 0.62 of five Earth regions for the linear models which indicates how much of the variation of a dependent variable ($M_W$) is explained by an independent variable ($m_b$) in a regression model. The lower value of $R$−square = 0.22 for the Europe region might suggest that the linear relationship between $M_W$ and $m_b > 6.1$ for this region is not the perfect model to explain their relationship. One of the suggested solutions for this issue is to apply non-linear models such as quadratic relationships or exponential models to express the relationship between dependent and independent magnitudes (Gasperini et al., 2013; Storchak et al., 2013; Sawires et al., 2019).

Table 2: The goodness of fit measures ($R$-squared, and $p$-value) for a linear fit of $M_W$ and $m_b \leq 6.1$ (a) and $m_b > 6.1$ (b) via the MCMC approach, for **Asia**-Oceania, Europe, Africa, North America, and South America, respectively.

a: For $m_b \leq 6.1$

| Region | Equation | R-squared | p-value |
|---|---|---|---|
| **Asia**-Oceania | $M_W = 0.923^{+0.090}_{-0.062} m_b + 0.576^{+0.475}_{-0.326}$ | 0.5561<br>0.5608 | 0.2096<br>0.1083 |
| Europe | $M_W = 0.922^{+0.093}_{-0.062} m_b + 0.581^{+0.478}_{-0.329}$ | 0.4200<br>0.4187 | 0.3099<br>0.2962 |
| Africa | $M_W = 0.811^{+0.088}_{-0.060} m_b + 1.267^{+0.458}_{-0.319}$ | 0.5076<br>0.5011 | 0.1970<br>0.1932 |
| North America | $M_W = 0.820^{+0.095}_{-0.065} m_b + 1.185^{+0.504}_{-0.349}$ | 0.4396<br>0.4371 | 0.1059<br>0.1007 |
| South America | $M_W = 0.864^{+0.093}_{-0.064} m_b + 0.884^{+0.505}_{-0.341}$ | 0.6235<br>0.6119 | 0.3156<br>0.3094 |

b: For $m_b > 6.1$

| Region | Equation | R-squared | p-value |
|---|---|---|---|
| **Asia**-Oceania | $M_W = 1.253^{+0.553}_{-0.371} m_b - 1.162^{+3.485}_{-2.370}$ | 0.4770<br>0.4712 | 0.1187<br>0.1174 |
| Europe | $M_W = 1.251^{+0.548}_{-0.374} m_b - 1.153^{+3.351}_{-0.058}$ | 0.2373<br>0.2297 | 0.3144<br>0.3158 |
| Africa | $M_W = 1.487^{+0.828}_{-0.584} m_b - 2.570^{+5.424}_{-3.710}$ | 0.2673<br>0.2690 | 0.3205<br>0.3156 |
| North America | $M_W = 1.370^{+0.797}_{-0.551} m_b - 1.892^{+5.103}_{-3.451}$ | 0.2668<br>0.2647 | 0.1041<br>0.1002 |
| South America | $M_W = 1.565^{+0.593}_{-0.412} m_b - 3.236^{+3.825}_{-2.630}$ | 0.4980<br>0.4962 | 0.2021<br>0.1451 |

Figure 6 exhibits the linear relationship of $M_W$ with $M_S \leq 6.1$ (red line) (and $M_S > 6.1$ (blue line)) in five Earth regions. We observe that the true value of the slope ranges from 0.566 to 0.619 for the small magnitude regime ($M_S \leq 6.1$) and 0.877 - 0.971 for large magnitudes ($M_S > 6.1$).

Based on Figure 6 and observing the large number of earthquake data and the high magnitude of seismic activity, there is high seismic activity in the middle of the eastern region. It is partly influenced by continental convergence and the shortening of the active crust between the African, Arabian, and Indian plates to the north concerning the Eurasian plate.

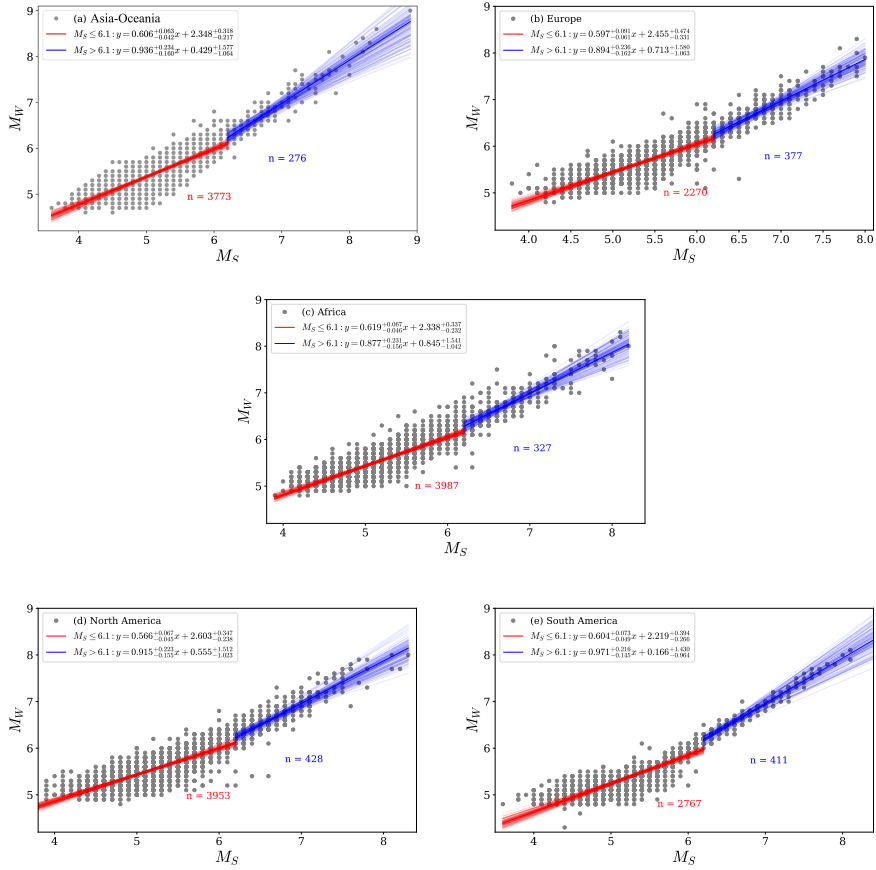

Figure 6: Transition relation between $M_W$ and $M_S$ using MCMC method for (a) **Asia**-Oceania, (b) Europe, (c) Africa, (d) North America, and (e) South America, respectively. *n* indicates the total of earthquakes for each region.

Table 3 tabulates the transition relationships between $M_W$ and $M_S$ for small earthquake magnitudes $M_S \leq 6.1$ and large magnitudes $M_S > 6.1$ for five Earth regions. The slope and intercept include the true values and upper and lower limits for linear relationships. The $p$−values more significant than the threshold 0.05 indicate that the linear relationships between $M_S$ and $M_W$ (null hypothesis) for all five Earth regions are not rejected. Interestingly, the $R$−square values are more significant than 0.6 validating the linear relationships between $M_W$ and $M_S$ for all five regions.

We obtained the true value of slope for small magnitude regime ranges 0.565 - 0.619 for $M_S \leq 6.1$ and for large magnitude regime varies from 0.877 to 0.971, for five Earth regions.

Scordilis (2006) investigated the transition relation of $M_W$ in terms of two regimes of $M_S$ and obtained the slopes of the linear relationship 0.67 ($3.0 \leq M_S \leq 6.1$), and 0.99 ($6.2 \leq MS \leq 8.2$). Our result for the same two regimes covers the Scordilis (2006) results. Also, KADİRİOĞLU and Kartal (2016) explored the relationship between $M_W$ and $M_S$ in two different regimes in the Turkey region with the orthogonal regression (OR) and ordinary least squares (OLS) methods. The slopes of the linear relationship lines were 0.5716 (OR) and 0.6524 (OLS) for $M_S \leq 5.4$ and 0.8126 (OR) and 0.7905 (OLS) for $M_S \geq 5.4$. The slopes of a linear relationship in the present work for small and large regimes of $M_S$ vary from 0.606 to 0.936 for **Asia**-Oceania. Lolli et al. (2014) investigated the conversion between teleseismic magnitudes ($M_S$ and $m_b$) from the Seismological Bulletin of the International Seismological Centre and moment magnitudes ($M_W$) from online moment tensor catalogs. They utilized the $\chi^2$ general orthogonal regression method, which accounts for measurement errors in both the predictor and response variables, as opposed to the ordinary least-squares regression method. Their results indicate that the slope of the $M_W$-$M_S$ relationship at lower magnitudes closely matches the theoretical value of 2/3 for $M_S < 5.0$. On the other hand, the slope of the $M_W$-$m_b$ relationship at higher magnitudes is expected to approach the theoretical value of 2, as proposed by Kanamori's model (Kanamori and Anderson, 1975), only near the upper limit of $m_b = 7.2$.

Boudebouda et al. (2024) studied the relationships for converting different magnitudes to the moment magnitude ($M_W$, GCMT) worked on the moment magnitudes data used for northern Algeria have been taken principally from the GCMT catalog and enhanced with the European-Mediterranean Regional Centroid Moment to estimate the magnitude scales tested against $M_W$ are the surface and body wave magnitudes released from the international seismological sources of ISC and NEIC for the same boundaries. They obtained the slope of $M_S$ and $M_W$ linear relation about 0.71 which is in the range of our finding for the slope obtained for Africa.

The differences between the findings of the present study and previous investigations may be due to some factors such as differences in the Earth region under study, period of recorded earthquakes, data centers, and applied algorithms for determining the parameters of the linear relationship between $M_W$ and $m_b$. Indeed, in the present work, we used the MCMC approach that applies the standard errors of $M_W$ calculated from the moment tensor using the error propagation method. However, the previous studies obtained the parameters of the linear relationship between magnitudes mainly based on the least square approach.

Table 3: The goodness of fit measures ($R$−square, and $p$−value) for a linear fit of $M_W$ and $M_S \leq 6.1$ (a) and $M_S > 6.1$ (b) via the MCMC approach, for **Asia**-Oceania, Europe, Africa, North America, and South America, respectively.

a: For $M_S \leq 6.1$

| Region | Equation | $R$-squared | $p$-value |
|---|---|---|---|
| **Asia**-Oceania | $M_W = 0.606^{+0.063}_{-0.042} M_S + 2.348^{+0.318}_{-0.217}$ | 0.7408 / 0.7341 | 0.3672 / 0.3314 |
| Europe | $M_W = 0.597^{+0.091}_{-0.061} M_S + 2.455^{+0.474}_{-0.331}$ | 0.6702 / 0.6676 | 0.2037 / 0.1964 |
| Africa | $M_W = 0.619^{+0.067}_{-0.046} M_S + 2.338^{+0.337}_{-0.232}$ | 0.7413 / 0.7395 | 0.4001 / 0.3825 |
| North America | $M_W = 0.566^{+0.067}_{-0.045} M_S + 2.603^{+0.347}_{-0.238}$ | 0.6651 / 0.6470 | 0.1154 / 0.1088 |
| South America | $M_W = 0.604^{+0.073}_{-0.049} M_S + 2.219^{+0.394}_{-0.266}$ | 0.7890 / 0.7823 | 0.1954 / 0.1837 |

b: For $M_S > 6.1$

| Region | Equation | $R$-squared | $p$-value |
|---|---|---|---|
| **Asia**-Oceania | $M_W = 0.936^{+0.234}_{-0.160} M_S + 0.429^{+1.577}_{-1.064}$ | 0.8940 / 0.8901 | 0.1045 / 0.1031 |
| Europe | $M_W = 0.894^{+0.236}_{-0.162} M_S + 0.713^{+1.580}_{-1.063}$ | 0.7994 / 0.7951 | 0.4940 / 0.4651 |
| Africa | $M_W = 0.877^{+0.231}_{-0.156} M_S + 0.845^{+1.541}_{-1.042}$ | 0.7519 / 0.7496 | 0.1457 / 0.1304 |
| North America | $M_W = 0.915^{+0.223}_{-0.155} M_S + 0.555^{+1.512}_{-1.023}$ | 0.7819 / 0.7750 | 0.2483 / 0.2471 |
| South America | $M_W = 0.971^{+0.216}_{-0.145} M_S + 0.166^{+1.430}_{-0.964}$ | 0.9753 / 0.9705 | 0.1357 / 0.1229 |

Additionally, we apply the MCMC approach to infer the parameters of the linear transition relationship between $M_W$ and $M_b$ as well $M_W$ and $M_S$ for global earthquakes, including $n$ =18,569 pairs of magnitudes. Figure 7 depicts the linear transition relationship between $M_W$ with $m_b$ (left panel) and $M_W$ with $M_S$ (right panel). For $m_b \leq 6.1$ with $n = 17,815$ (and $m_b > 6.1$ with $n = 754$), we obtain the $\alpha = 0.863$ and $\beta = 0.945$ ($\alpha = 1.374$ and $\beta = -1.924$ ). Also, for $M_S \leq 6.1$ with $n = 16750$ (and $M_S > 6.1$ with $n = 1819$ ), we obtain the $\alpha = 0.581$ and $\beta = 2.487$ ($\alpha = 0.921$ and $\beta = 0.521$ ). Our findings for the slope of $M_W$ with $M_b$ and $M_W$ with $M_S$ agree with the previous studies. Scordilis (2006) reported the slope 0.85 ± 0.04 for $M_W$ and $m_b \leq 6.2$ for 39,784 global earthquakes. They obtained $\alpha = 0.99 \pm 0.02$ and $\alpha = 0.08 \pm 0.13$ for the relationship of $M_W$ with $M_S \leq 6.2$ and $M_S > 6.2$, respectively. Di Giacomo et al. (2015) investigated the slope of linear relationship for $M_W$ with $m_b \leq 6.5$ 1.47 for 14,862 earthquakes and 0.67 for $M_W$ with $M_S \leq 8.8$ of 16,978 earthquakes. Das et al. (2018) determined the slope of the linear transition relationship for $M_W$ with $m_b$ 0.998 for 19,466 earthquake.

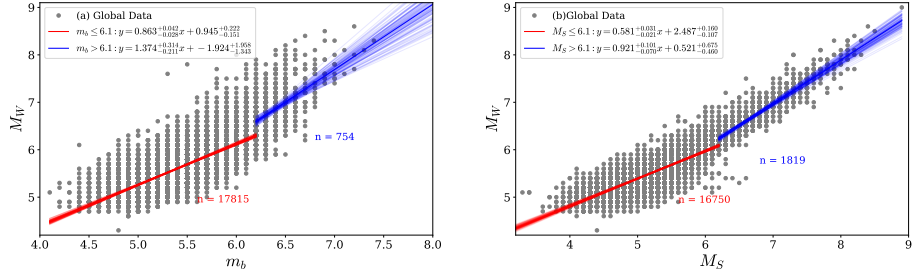

Figure 7: Linear transition relationship of $M_W$ with $m_b$ (left panel) and $M_W$ with $M_S$ (right panel) for magnitudes less than 6.1 (red line) and greater than 6.1 (blue line) for 18,569 earthquakes recorded by Harvard. $n$ indicates the number of earthquakes. We represent the slope ($\alpha$) and intercept ($\beta$) together with upper (superscript) and lower (subscript) limits of standard errors for the linear relationship $y = \alpha x + \beta$ for magnitudes $\leq 6.1$ and $> 6.1$.

For global data, the slopes and intercepts ($\alpha, \beta$) obtained using the MCMC approach are as follows:

- For $m_b \leq 6.1$:
$$M_W = 0.863^{+0.042}_{-0.028}\, m_b + 0.945^{+0.222}_{-0.151}$$

- For $m_b > 6.1$:
$$M_W = 1.374^{+0.314}_{-0.211}\, m_b - 1.924^{+0.965}_{-1.343}$$

- For $M_S \leq 6.1$:
$$M_W = 0.581^{+0.031}_{-0.021}\, M_S + 2.487^{+0.160}_{-0.107}$$

- For $M_S > 6.1$:
$$M_W = 0.921^{+0.101}_{-0.070}\, M_S + 0.521^{+0.675}_{-0.460}$$

# 6 Conclusion

Estimating the accurate parameters of the relationship of various earthquake magnitudes is essential in seismological studies. Since the $m_b$ and $M_S$ quickly saturate, they are not suitable parameters to accurately estimate the size of an earthquake and the risks involved. This paper investigated empirical relationships between earthquake magnitudes expressed on a large scale equivalent to moments. Due to the statistical significance of regional differences for the moment ratio (e.g., $M_S/m_b$) and different source mechanisms of earthquakes, we grouped the earthquake magnitudes for five Earth regions, namely **Asia**-Oceania, Europe, Africa, North America, and South America (Figure 1). The resultant relationships may provide a tool for compiling homogeneous earthquake catalogs to estimate magnitudes and uncertainties. We showed that the probability density function of magnitude errors deviates from the normal distribution. Therefore, the list square fitting approaches are not adequate to drive the transition relationship between different magnitudes. So, we used the MCMC approach to obtain the parameters of the straight-line models as the relationship of magnitudes ($m_b$, $M_W$) and ($M_S$, $M_W$).

Applying the error propagation method, we calculated the standard errors for $M_W$ that are related to the moment tensor (Equations 4-6). The distribution of standard error for $M_W$ deviates from the Gaussian distribution (Figure 2), which recommended applying the MCMC approach to determine the parameters of a linear model for $M_W$ and $m_b$ (or $M_S$) instead of the least square method. In two regimes of magnitudes $\leq 6.1$ and $> 6.1$ for both $m_b$ and $M_S$ of each Earth region, we obtained the slope and intercept of the linear transition relationships with $M_W$ (Figures 5 and 6). The hypothesis testing $p-$value significantly greater than threshold 0.05 indicated that these linear relationships (null hypothesis) between earthquake magnitudes are not rejected. However, the lower $R-$square value for some regions suggests a nonlinear relationship for magnitudes (Tables 2 and 3). Applying the MCMC method for 18,569 global earthquakes gives the slope for $m_b \leq 6.1$ ($m_b > 6.1$) 0.862 (1.377), while we find the slope for $M_S \leq 6.1$ ($M_S > 6.1$) 0.581 (0.921).

Finally, the adequate transition relation of moment, surface, and body magnitudes obtained here may help study the physics source of earthquakes in different Earth regions. We conclude that correlations of different magnitudes are slightly different for different Earth regions that may related to their various physical sources. The relationships of different magnitude moments help monitor the earthquakes, which may observed at one or two-moment magnitudes at some stations. The relationship of various magnitudes is useful for monitoring and predicting earthquake magnitudes.

# 7 Data availability

The dataset that supports the findings of this study is available from the authors upon request [zahra.firoozeh@znu.ac.ir and safari@znu.ac.ir].
The earthquake data supporting this study's findings are available at the following link: `https://www.globalcmt.org/CMTsearch.html`.

# 8 Acknowledgments

We acknowledge HRVD at `https://www.globalcmt.org/CMTsearch.html` data center for making the data publicly available. We use the Python code of (`https://emcee.readthedocs.io/en/stable`) in some fitting. We sincerely appreciate the Reviewers' valuable comments and suggestions, which helped us to improve the manuscript's quality.

# 9 Declarations

Conflict of interest The authors declare that they have no known competing interests.